\documentclass[amstex,amsfonts,12pt,thmsa,sw20bams]{article}

\input tcilatex
\begin{document}

\author{Choulakian, V. \and Universit\'{e} de Moncton, Canada \and email:
vartan.choulakian@umoncton.ca; }
\title{Matrix Factorizations Based on Induced Norms}
\date{August 2015}
\maketitle

\begin{abstract}
We decompose a matrix ${\bf Y}$ into a sum of bilinear terms in a stepwise
manner, by considering ${\bf Y}${\bf \ }as a mapping from the finite
dimensional space $l_{r}^{n}$ to the space $l_{p}^{m}$. We provide
transition formulas, and represent them in a duality diagram, thus
generalizing the well known duality diagram in the french school of data
analysis.\ As an application, we introduce a family of Euclidean
multidimensional scaling models.


Classification AMS 2010: 62H25
\end{abstract}

\section{Introduction}

Matrix factorization, named also decomposition, in data analysis is at the
core of factor analysis; and one of its principal aims, as clearly stated by
Hubert et al. (2000), is to visualize geometrically the statistical
association existing among the rows or the columns of the matrix. So the way
that we factorize a matrix is of fundamental interest and concern in
statistics. What is surprising is that the oldest method, the centroid
factorization, see Burt (1917) and Thurstone (1931), has been rediscovered
recently many times, see for instance proposal 1 in McCoy and Tropp (2011).
Singular value decomposition (SVD) is the most used matrix decomposition
method in statistics; the aim of this paper is to present in a coherent way
the theory of SVD-like matrix factorizations based on subordinate or induced
norms; and at the same time, review the existing literature. This
presentation generalizes the SVD by embedding it in a larger family: It
belongs to the class of optimal biconjugate decompositions; biconjugate
decompositions are based on Wedderburn rank-one reduction theorem as
described by Chu et al. (1995). Other alternative generalization of SVD,
GSVD, is presented by Hubert et al. (2000), and which forms the basis of the
french school of data analysis as reviewed recently by Holmes (2008) and De
La Cruz and Holmes (2011). We also incorporate the GSVD in our
representation.

This paper is organized as follows: Section 2 presents the preliminaries
concerning induced or subordinate matrix norms; section 3 presents the
matrix factorizations based on induced norms, and we conclude in section 4.

\section{Preliminaries on real Banach spaces $l_{p}^{n}$}

We start with some preliminaries and at the same time introduce notation. We
note:

$l_{p}^{n}:=(%
\textbf{R}
$ $^{n},\ ||.||_{p})$ is a finite dimensional Banach space; that is, $%
\textbf{R}
$ $^{n}$ is $n$-dimensional complete vector space with the $p$-norm, $%
||.||_{p},$ for $p\geq 1.$ For an ${\bf x}$ $\in $ $%
\textbf{R}
$ $^{n},$ its $p$-norm is defined as $||{\bf x}||_{1}=\sum_{i=1}^{n}|x_{i}|$
for $p=1,$ $||{\bf x}||_{p}=(\sum_{i=1}^{n}|x_{i}|^{p})^{1/p}$ for $p>1$,
and $||{\bf x}||_{\infty }=\max_{i=1}^{n}|x_{i}|$ for $p=\infty .$

The norm $||{\bf x}||_{p}$ has the following four properties\bigskip

(N1) $||{\bf x}||_{p}\geq 0$

(N2) $||{\bf x}||_{p}=0$ iff ${\bf x=0}$

(N3) $||\alpha {\bf x}||_{p}=\alpha ||{\bf x}||_{p}$ for $\alpha \in
\textbf{R}
$

(N4) $||{\bf x+y}||_{p}\leq ||{\bf x}||_{p}+||{\bf y}||_{p}\bigskip $

(N4) implies: $|||{\bf x}||_{p}-||{\bf y}||_{p}|\leq ||{\bf x-y}||_{p}$,
from which we deduce that the $p$-norm is a continuous mapping of $%
\textbf{R}
$ $^{n}$ into $%
\textbf{R}
.$

The proof of (N4) is based on H\"{o}lder and Minkowski inequalities.

We define the unit sphere to be
\[
S_{p}^{n}=\{{\bf x}\in
\textbf{R}^{\mbox{n}}:{|| {\bf x} ||}_{p} =\mbox{1}\},
\]%
and $(p,p_{1})$ designate the conjugate pair, that is, $\frac{1}{p}+\frac{1}{%
p_{1}}=1$ for $p\geq 1$ and $p_{1}\geq 1.\bigskip $

{\bf H\"{o}lder inequality}:
\[
<{\bf x}^{\ast },{\bf x>\ \leq \ }||{\bf x}^{\ast }||_{p_{1}}||{\bf x}%
||_{p}\ \mbox{\ \ for \ }{\bf x}^{\ast }\in l_{p_{1}}^{n}\mbox{ \ \ and \ \ }%
{\bf x}\in l_{p}^{n}
\]%
or
\[
<{\bf x}^{\ast },{\bf x>\ \leq \ }||{\bf x}||_{p}\ \mbox{\ \ for \ }{\bf x}%
^{\ast }\in S_{p_{1}}^{n}\mbox{ \ \ and \ \ }{\bf x}\in l_{p}^{n},
\]%
or%
\[
<{\bf x}^{\ast },{\bf x>\ \leq \ }1\ \mbox{\ \ for \ }{\bf x}^{\ast }\in
S_{p_{1}}^{n}\mbox{ \ \ and \ \ }{\bf x}\in S_{p}^{n}.
\]

Note that $<{\bf x}^{\ast },{\bf x>\ =\ }\sum_{i=1}^{n}x_{i}^{\ast }x_{i}=(%
{\bf x}^{\ast })^{\prime }{\bf x=x}^{\prime }{\bf x}^{\ast }$, where ${\bf x}%
^{\prime }$ is the transpose of the row vector ${\bf x;}$ further, $<{\bf x}%
^{\ast },{\bf x>}$ represents a scalar product only when the conjugate pair $%
(p,p_{1})=(2,2).$ The next result is an application of H\"{o}lder
inequality.\bigskip

{\bf Lemma 1}: Let ${\bf x}$ $\in $ $l_{p}^{n},$ then there exists a norming
functional $\varphi ({\bf x})\in S_{p_{1}}^{n}$ such that $<\varphi ({\bf x}%
),{\bf x>\ =\ }||{\bf x}||_{p}=\sup_{{\bf x}^{\ast }}<{\bf x}^{\ast },{\bf x>%
}$ $\ \ $subject to $\ {\bf x}^{\ast }\in S_{p_{1}}^{n}.$ Explicitly we have:

\begin{eqnarray*}
\varphi ({\bf x}) &=&(v_{j}=sgn(x_{j})\ )\ \ \ \ \ \ \ \mbox{for\ \ \ }p=1 \\
&=&(v_{j}=sgn(x_{j})\ |\frac{x_{j}}{||{\bf x}||_{p}}|^{p-1})\ \ \ \ \ \ \
\mbox{for\ \ \ }p>1 \\
&=&{\bf e}_{\alpha }\mbox{ }sgn(x_{\alpha })\ \mbox{ \ \ \ \ \ \ \ \ for\ \
\ \ }p=\infty ,
\end{eqnarray*}%
where \{${\bf e}_{\beta }:\beta =1,...,n\}$ designates the canonical basis
and $x_{\alpha }=\arg \max_{\beta =1}^{n}|x_{\beta }|$.\bigskip

Remark: In more general settings, Lemma 1 is proven as a corollary to the
famous Hahn-Banach theorem, see for instance Kreyszig (1978, p.223).\bigskip

{\bf Example 1}: Consider the vector ${\bf x}^{^{\prime }}=(1\ \ \ 2\ \ \
-1\ \ \ -2).$

a) If ${\bf x}$ $\in $ $l_{2}^{4},$ then $||{\bf x}||_{2}=10^{1/2}$ \ and \ $%
\varphi ({\bf x})=\frac{{\bf x}}{10^{1/2}}\in S_{2}^{4}$ and $<\varphi ({\bf %
x}),{\bf x>\ =\ }10^{1/2}.$ Explicitly $\varphi ({\bf x})^{\prime }=(1\ \ \
2\ \ \ -1\ \ \ -2)/10^{1/2}.$

b) If ${\bf x}$ $\in $ $l_{1}^{4},$ then $||{\bf x}||_{1}=6$ \ and \ $%
\varphi ({\bf x})=sgn({\bf x})\in S_{\infty }^{4}$ and $<\varphi ({\bf x}),%
{\bf x>\ =\ }6.$ Explicitly $\varphi ({\bf x})^{\prime }=(1\ \ \ 1\ \ \ -1\
\ \ -1).$

c) If ${\bf x}$ $\in $ $l_{\infty }^{4},$ then $||{\bf x}||_{\infty }=2$ \
and \ $\varphi ({\bf x})=-{\bf e}_{4}\in S_{1}^{4}$ and $<\varphi ({\bf x}),%
{\bf x>\ =\ }2.$ Explicitly $\varphi ({\bf x})^{\prime }=(0\ \ \ 0\ \ \ 0\ \
\ -1).$ Another value is: $\varphi ({\bf x})^{\prime }=(0\ \ \ 1\ \ \ 0\ \ \
0).$

d) If ${\bf x}$ $\in $ $l_{3}^{4},$ then $||{\bf x}||_{3}=18^{1/3}$ \ and \ $%
\varphi ({\bf x})=(v_{j}=\frac{x_{j}^{2}}{18^{2/3}}sgn(x_{j}))\in
S_{1.5}^{4} $ and $<\varphi ({\bf x}),{\bf x>\ =\ }18^{1/3}.$ Explicitly $%
\varphi ({\bf x})^{\prime }=(1\ \ \ 4\ \ \ -1\ \ \ -4)/18^{2/3}.\bigskip $

Let $B(l_{r}^{n},l_{p}^{m})$ be the set of bounded linear maps (operators)
from $l_{r}^{n}$ to $l_{p}^{m},$ which we identify with the set of $m\times
n $ real matrices in the usual way. If ${\bf A}\in B(l_{r}^{n},l_{p}^{m})$,
then ${\bf A}^{\prime }\in B(l_{p_{1}}^{m},l_{r_{1}}^{n}).$ Let ${\bf A}\in
B(l_{r}^{n},l_{p}^{m})$ be an operator, then its induced or subordinate norm
is defined to be%
\begin{equation}
||{\bf A}||_{r\rightarrow p}=\max \{||{\bf Au}||_{p}:{\bf u}\in S_{r}^{n}\}.
\end{equation}%
Then
\begin{equation}
||{\bf A}^{\prime }||_{p_{1}\rightarrow r_{1}}=\max \{||{\bf A}^{\prime }%
{\bf v}||_{r_{1}}:{\bf v}\in S_{p_{1}}^{m}\},
\end{equation}%
and the next theorem is a well known central result.\bigskip

{\bf Theorem 1}:
\begin{eqnarray*}
||{\bf A}||_{r\rightarrow p} &=&||{\bf A}^{\prime }||_{p_{1}\rightarrow
r_{1}} \\
&=&\max \{{\bf v}^{\prime }{\bf Au}:{\bf u}\in S_{r}^{n}\mbox{ \ \ and \ \ }%
{\bf v}\in S_{p_{1}}^{m}\} \\
&=&{\bf v}_{1}^{\prime }{\bf Au}_{1}\mbox{ \ \ for \ \ }{\bf u}_{1}\in
S_{r}^{n}\mbox{ \ \ and \ \ }{\bf v}_{1}\in S_{p_{1}}^{m} \\
&=&{\bf v}_{1}^{\prime }{\bf a}_{1}={\bf b}_{1}^{\prime }{\bf u}_{1}=\lambda
_{1}
\end{eqnarray*}

where
\begin{equation}
{\bf Au}_{1}={\bf a}_{1}\mbox{ \ and \ }{\bf v}_{1}=\varphi ({\bf a}_{1})
\end{equation}%
and%
\begin{equation}
{\bf A}^{\prime }{\bf v}_{1}{\bf =b}_{1}\mbox{ \ and \ }{\bf u}_{1}=\varphi (%
{\bf b}_{1}),
\end{equation}%
and the last two equations are known as transition formulas.\bigskip

{\it Proof}: For ${\bf u}\in S_{r}^{n}$ \ \ and \ \ ${\bf v}\in
S_{p_{1}}^{m},$ we consider the bilinear form
\begin{eqnarray}
\lambda ({\bf u},{\bf v}) &=&{\bf v}^{\prime }{\bf Au}  \nonumber \\
&\leq &||{\bf Au}||_{p}\mbox{ \ by \ H\"{o}lder inequality for }{\bf Au}\in
l_{p}^{n}  \nonumber \\
&\leq &\max_{{\bf u}\in S_{r}^{n}}||{\bf Au}||_{p}=||{\bf A}||_{r\rightarrow
p}\mbox{ \ by (1)} \\
&=&||{\bf Au}_{1}||_{p}={\bf v}_{1}^{\prime }{\bf Au}_{1}\mbox{ \ where }%
{\bf v}_{1}=\varphi ({\bf Au}_{1})=\varphi ({\bf a}_{1}) \\
&=&\max_{{\bf v}\in S_{p_{1}}^{m}}{\bf v}^{\prime }{\bf Au}_{1}\mbox{ \ \ by
\ Lemma 1}  \nonumber \\
&=&\max_{{\bf v}\in S_{p_{1}}^{m}}\max_{{\bf u}\in S_{r}^{n}}{\bf v}^{\prime
}{\bf Au.}  \nonumber
\end{eqnarray}

\bigskip Now using (5 and 6) and replacing ${\bf A}$ by ${\bf A}^{\prime }$
we have%
\begin{eqnarray*}
||{\bf A}||_{r\rightarrow p} &=&{\bf v}_{1}^{\prime }{\bf Au}_{1} \\
&=&{\bf u}_{1}^{\prime }{\bf A}^{\prime }{\bf v}_{1}\mbox{ \ where }{\bf u}%
_{1}=\varphi ({\bf A}^{\prime }{\bf v}_{1})=\varphi ({\bf b}_{1}) \\
&=&||{\bf A}^{\prime }||_{p_{1}\rightarrow r_{1}},
\end{eqnarray*}%
which is the required result.\bigskip

The transition formulas (3 and 4) can be represented by the following
duality diagram\bigskip

\ \ \ \ \ \ \ \ \ \ \ \ \ \ ${\bf A}$

\ \ \ \ \ $S_{r}^{n}\ \ \ \longrightarrow \ \ l_{p}^{m}$

$\ {\bf \ }\varphi \uparrow \ \ \ \ \ \ \ \ \ \ \ \downarrow \varphi $

\ \ \ \ \ $l_{r_{1}}^{n}\ \ \longleftarrow \ \ S_{p_{1}}^{m}\ $

\ \ \ \ \ \ \ \ \ \ \ \ \ $\ {\bf A}^{\prime }\bigskip $

{\bf Remark 1}:

a) The geometrical-statistical interpretation of Theorem 1 is that $\lambda
_{1}$ is the largest dispersion value by which the operator ${\bf A}$
stretches an element of ${\bf u}\in S_{r}^{n}$; ${\bf u}_{1}$ is called the
first principal axis of the rows of ${\bf A}${\bf , }and{\bf \ }${\bf a}_{1}$
represents the projected values of the rows of ${\bf A}$ on ${\bf u}_{1}$,
and we name it the first projected row factor or the first principal
component. And by duality, we also have $\lambda _{1}$ is the largest
dispersion value by which the operator ${\bf A}^{\prime }$ stretches an
element of ${\bf v}\in S_{p_{1}}^{m}$; ${\bf v}_{1}$ is called the first
principal axis of the columns of ${\bf A}${\bf , }and{\bf \ }${\bf b}_{1}$
is the first column projected factor which represents the projected values
of the columns of ${\bf A}$ on ${\bf v}_{1}.$ Essentially, we are computing
the quintiplet $({\bf a}_{1},{\bf b}_{1},{\bf u}_{1},{\bf v}_{1},\lambda
_{1}).$

b) The vectors ${\bf a}_{1},{\bf b}_{1},{\bf u}_{1}$ and ${\bf v}_{1}$
belong to four different spaces: ${\bf a}_{1}\in l_{p}^{m},$ $\ {\bf b}%
_{1}\in l_{r_{1}}^{n},$ ${\bf u}_{1}\in S_{r}^{n}\subset l_{r}^{n}$ \ \ and
\ \ ${\bf v}_{1}\in S_{p_{1}}^{m}\subset l_{p_{1}}^{m}.$

c) The transition formulas provide us an iterative algorithm to compute a
maximum of $\{||{\bf Au}||_{p}:{\bf u}\in S_{r}^{n}\};$ this maximum value
can be a relative maximum. The norm $||{\bf A}||_{r\rightarrow p}$
corresponds to the absolute maximum. The algorithm is named the power method
for $l_{p}$ norm by Boyd (1974); Wold's (1966) NIPALS (nonlinear iterative
partial alternating least squares) algorithm, named also criss-cross
regression by Gabriel and Zamir (1979), is a particular case. The algorithm
can be summarized in the following way, where ${\bf b}$ is a starting
value:\bigskip

Step 1: ${\bf u=}\varphi {\bf (b)}$, ${\bf a=Au}$\ and $\lambda ({\bf a)}%
=\left\vert \left\vert {\bf a}\right\vert \right\vert _{p};$

Step 2: ${\bf v=}\varphi {\bf (a),}$ ${\bf b=A}^{\prime }{\bf v}$\ and $%
\lambda ({\bf b)}=\left\vert \left\vert {\bf b}\right\vert \right\vert
_{r_{1}};$

Step 3: If $\lambda ({\bf b)-}\lambda ({\bf a)>}0{\bf ,}$ go to Step 1;
otherwise, stop.\bigskip

The proof of the convergence of the algorithm is based on application of H%
\"{o}lder inequality twice: Let ${\bf u}^{(k)},$ ${\bf v}^{(k)}$ and $%
\lambda ^{(k)}$ for $k\geq 1$ represent the $k$th iteration values, then:%
\begin{eqnarray*}
\lambda ^{(k)} &=&{\bf v}^{(k)\prime }{\bf Au}^{(k)} \\
&\leq &({\bf v}^{(k)\prime }{\bf A)}\varphi {\bf (A}^{\prime }{\bf v}^{(k)})%
\mbox{ \ \ by H\"{o}lder \ inequality} \\
&=&{\bf v}^{(k)\prime }{\bf Au}^{(k+1)} \\
&\leq &\varphi ({\bf Au}^{(k+1)})^{\prime }({\bf Au}^{(k+1)})\mbox{ \ \ by H%
\"{o}lder \ inequality} \\
&=&{\bf v}^{(k+1)\prime }{\bf Au}^{(k+1)}=\lambda ^{(k+1)}.
\end{eqnarray*}

The rows or the columns of ${\bf A}$ can be used as starting values for $%
{\bf a}$ or ${\bf b}$.

\subsection{Particular norms}

Let ${\bf A}\in B(l_{r}^{n},l_{p}^{m})$, then ${\bf A}^{\prime }\in
B(l_{p_{1}}^{m},l_{r_{1}}^{n}).$ In general the conjugate pairs $(r,r_{1})$
and $(p_{1},p)$ are not equal, which implies that the geometry of the rows
is different from the geometry of the columns. If $(r,r_{1})=(p_{1},p)=(r,p)$%
, then the geometric structure defined on the rows of ${\bf A}$ is identical
to the geometric structure defined on the columns of ${\bf A;}$ this class
was named transposition invariant by Choulakian (2005).

For two particular values of the conjugate pairs $(r,r_{1}),$ explicit
formulas are available; however, the spectral norm is the most well known,
which is transposition invariant.

a) $(r,r_{1})=(\infty ,1),$ then%
\begin{eqnarray*}
||{\bf A}||_{\infty \rightarrow p} &=&||{\bf A}^{\prime
}||_{p_{1}\rightarrow 1}\mbox{ \ by Thorem 1} \\
&=&\max_{{\bf u}}{\bf ||Au||}_{p}\ \ \mbox{subject to }{\bf u}\in \left\{
-1,+1\right\} ^{n}.
\end{eqnarray*}

The proof is based on H\"{o}lder inequality: For any ${\bf v}\in
S_{p_{1}}^{m}$ consider
\begin{eqnarray*}
||{\bf A}^{\prime }{\bf v}||_{1} &=&{\bf u}^{\prime }{\bf A}^{\prime }{\bf v}%
\mbox{ \ \ for \ }{\bf u=}sgn{\bf (A}^{\prime }{\bf v)}\in \left\{
-1,+1\right\} ^{n}\mbox{ \ by \ Lemma 1}, \\
&\leq &{\bf ||Au||}_{p}\ \ \ \mbox{for \ }{\bf u}\in \left\{ -1,+1\right\}
^{n}\mbox{ \ by \ H\"{o}lder inequality,} \\
&\leq &\max {\bf ||Au||}_{p}\ \ \ \mbox{for \ }{\bf u}\in \left\{
-1,+1\right\} ^{n} \\
&=&{\bf ||Au}_{1}{\bf ||}_{p}\mbox{ \ where }{\bf u}_{1}=\arg \max {\bf %
||Au||}_{p}\mbox{ \ subject to }{\bf u}\in \left\{ -1,+1\right\} ^{n}.
\end{eqnarray*}%
By Lemma 1, if ${\bf v=}\varphi ({\bf Au}_{1})$, then $||{\bf A}^{\prime
}||_{p_{1}\rightarrow 1}=\max \{||{\bf A}^{\prime }{\bf v}||_{1}:{\bf v}\in
S_{p_{1}}^{m}\}=\max_{{\bf u}}{\bf ||Au||}_{p}\ \ $subject to ${\bf u}\in
\left\{ -1,+1\right\} ^{n}$, which is the required result.

b) $(r,r_{1})=(1,\infty ),$ then%
\begin{eqnarray*}
||{\bf A}||_{1\rightarrow p} &=&||{\bf A}^{\prime }||_{p_{1}\rightarrow
\infty }\mbox{ \ by Theorem 1} \\
&=&\max_{\alpha =1}^{m}||{\bf A}_{\ast \alpha }||_{p},
\end{eqnarray*}%
where ${\bf A}_{\ast \alpha }$ is the $\alpha $th column of ${\bf A}$.

The proof is similar to the proof in a). For any ${\bf v}\in S_{p_{1}}^{m}$
consider
\begin{eqnarray*}
||{\bf A}^{\prime }{\bf v}||_{\infty } &=&{\bf u}^{\prime }{\bf A}^{\prime }%
{\bf v}\mbox{ \ \ for \ }{\bf u=e}_{\alpha }\mbox{ }sgn(x_{\alpha })\ \
\mbox{and \ \ }x_{\alpha }=\arg \max_{\beta =1}^{n}|{\bf A}^{\prime }{\bf v}|%
\mbox{ \ by Lemma 1}, \\
&\leq &{\bf ||Au||}_{p}\ \ \ \mbox{for \ }{\bf u=e}_{\alpha }\mbox{ }%
sgn(x_{\alpha })\ \ \mbox{and \ \ }x_{\alpha }=\arg \max_{\beta =1}^{n}|{\bf %
A}^{\prime }{\bf v}|\mbox{ \ by \ H\"{o}lder inequality,} \\
&\leq &{\bf ||Au}_{1}{\bf ||}_{p}\mbox{ \ \ for }{\bf u}_{1}=\arg \max_{{\bf %
u}}{\bf ||Ae_{\alpha }||}_{p}, \\
&=&\max_{\alpha =1}^{m}||{\bf A}_{\ast \alpha }||_{p}.
\end{eqnarray*}%
By Theorem 1, if ${\bf v=}\varphi ({\bf Au}_{1})$, then $\max \{||{\bf A}%
^{\prime }{\bf v}||_{\infty }:{\bf v}\in S_{p_{1}}^{m}\}=$ $\max_{\alpha
=1}^{m}||{\bf A}_{\ast \alpha }||_{p}=$ $\max_{{\bf u}}{\bf ||Au||}_{p}\ \ $%
subject to ${\bf u=e}_{\alpha }$ for $\alpha =1,...,n,$ which is the
required result.

c) For $(r,r_{1})=(2,2),$ then
\begin{eqnarray*}
||{\bf A}||_{2\rightarrow 2} &=&||{\bf A}^{\prime }||_{2\rightarrow 2} \\
&=&\sqrt{\lambda _{\max }({\bf AA}^{\prime })} \\
&=&\sqrt{\lambda _{\max }({\bf A}^{\prime }{\bf A})},
\end{eqnarray*}%
where $\lambda _{\max }$ is the greatest eigenvalue of ${\bf AA}^{\prime }$
or ${\bf A}^{\prime }{\bf A,}$ and it is named spectral norm.

Drakakis and Pearlmutter (2009) and Lewis (2010) discuss the following nine
cases of $||{\bf A}||_{r\rightarrow p}$ for $r,p=1,2,3,$ which can be easily
deduced from the above results.

\section{Matrix factorizations}

Let ${\bf X}\in B(l_{r}^{n},l_{p}^{m}).$ Let $({\bf a}_{1},{\bf b}_{1})$ be
the first projected factors associated with $\lambda _{1}.$ We repeat the
above procedure on the residual dataset
\begin{eqnarray}
{\bf X}^{(1)} &{\bf =}&{\bf X-a}_{1}{\bf b}_{1}^{\prime }/\lambda _{1} \\
&=&({\bf I}_{m}-{\bf P}_{a_{1}}){\bf X}  \nonumber \\
&=&{\bf X}({\bf I}_{n}-{\bf P}_{b_{1}}){\bf ,}  \nonumber
\end{eqnarray}%
where ${\bf P}_{{\bf a}_{1}}={\bf a}_{1}{\bf v}_{1}^{\prime }/\lambda _{1}$
is the projection operator on ${\bf a}_{1}\in l_{p}^{m}$, because ${\bf P}_{%
{\bf a}_{1}}^{2}={\bf P}_{{\bf a}_{1}}.$ Similarly, ${\bf P}_{{\bf b}_{1}}=%
{\bf b}_{1}{\bf u}_{1}^{\prime }/\lambda _{1}$ is the projection operator on
${\bf b}_{1}\in l_{r_{1}}^{n}$. We note that the $rank({\bf X}^{(1)}{\bf )=}$
$rank({\bf X)-}1,$ because
\begin{equation}
{\bf X}^{(1)}{\bf u}_{1}={\bf 0}\mbox{ \ and }{\bf X}^{(1)\prime }{\bf v}%
_{1}={\bf 0;}
\end{equation}%
which implies that
\begin{equation}
{\bf u}_{1}^{\prime }{\bf b}_{2}=0\mbox{ \ and \ }{\bf v}_{1}^{\prime }{\bf a%
}_{2}=0\mbox{ \ }.
\end{equation}%
Equations (7,8,9) are known as Wedderburn's rank one reduction formula, see
Chu, Funderlic and Golub (1995). By repeating the above procedure we get the
data reconstitution formula for the matrix ${\bf X}$ as a function of the
projected row and column factor coordinates $({\bf a}_{\alpha },{\bf b}%
_{\alpha })$ associated with the dispersion values $\lambda _{\alpha },$ for
$\alpha =1,...,k,$ and $k=rank({\bf X}),$

\begin{equation}
{\bf X=\sum_{\alpha =1}^{k}}\mbox{ }{\bf a}_{\alpha }{\bf b}_{\alpha
}^{\prime }/\lambda _{\alpha }
\end{equation}%
or elementwise%
\[
x_{ij}=\sum_{\alpha =1}^{k}a_{\alpha }(i)b_{\alpha }(j)/\lambda _{\alpha }.
\]%
Equation (10) represents the decomposition of ${\bf X}$ based on $%
l_{r}^{n}\rightarrow l_{p}^{m}$ induced norm.

\subsection{The case of X symmetric}

When the matrix ${\bf X}$ is symmetric, we can have a symmetric
decomposition or a nonsymmetric factorization.

a) If the norms are transposition invariant, that is, the conjugate pairs $\
(r,r_{1})=(p_{1},p)=(r,p)$, then

\[
{\bf X=\sum_{\alpha =1}^{k}}\mbox{ }{\bf a}_{\alpha }{\bf a}_{\alpha
}^{\prime }/\lambda _{\alpha },
\]%
for the geometric structure defined on the rows of ${\bf X}$ is identical to
the geometric structure defined on the columns of ${\bf X.}$

b) If the norms are not transposition invariant, that is, the conjugate
pairs $\ (r,r_{1})\ \neq (p_{1},p)$, then

\[
{\bf X=\sum_{\alpha =1}^{k}}\mbox{ }{\bf a}_{\alpha }{\bf b}_{\alpha
}^{\prime }/\lambda _{\alpha },
\]%
for the geometric structure defined on the rows of ${\bf X}$ is different
from the geometric structure defined on the columns of ${\bf X.}$

\subsection{A\ review}

Here, we review published discussed cases in the statistical literature.

a) The centroid decomposition based on $||{\bf A}||_{\infty \rightarrow 2}=||%
{\bf A}^{\prime }||_{2\rightarrow 1};$ its transition formulas are%
\[
{\bf Au}_{1}={\bf a}_{1}\mbox{ \ and \ }{\bf v}_{1}={\bf a}_{1}/\sqrt{{\bf a}%
_{1}^{\prime }{\bf a}_{1}}
\]%
and%
\[
{\bf A}^{\prime }{\bf v}_{1}{\bf =b}_{1}\mbox{ \ and \ }{\bf u}_{1}=sgn({\bf %
b}_{1});
\]%
it is the oldest to our knowledge. First used by Burt (1917), then by
Thurstone (1931) to factorize covariance matrices, and used extensively in
the psychometric literature before the advent of the computers, see for
instance Thurstone (1947), Horst (1965) and Harman (1967). Burt-Thurstone
formulation was based on the following criterion:
\begin{equation}
\max {\bf u}^{\prime }{\bf A}^{\prime }{\bf Au}\ \ \mbox{subject to }{\bf u}%
\in \left\{ -1,+1\right\} ^{n};
\end{equation}%
its relationship with the matrix norm formulation was shown by Choulakian
(2003). In different, but related contexts, it is discussed by Galpin and
Hawkins (1987), Chu and Funderlic (2002), Choulakian (2005), Kwak (2008),
McCoy and Tropp (2011). Further, Choulakian (2012) considered it as a
particular MAXBET procedure which takes into account the block structure of
the variables.

b) $||{\bf A}||_{1\rightarrow 1}=||{\bf A}^{\prime }||_{\infty \rightarrow
\infty }$ is used by Galpin and Hawkins (1987); its transition formulas are%
\[
{\bf Au}_{1}={\bf a}_{1}\mbox{ \ and \ }{\bf v}_{1}=sgn({\bf a}_{1})
\]%
and%
\[
{\bf A}^{\prime }{\bf v}_{1}{\bf =b}_{1}\mbox{ \ and \ }{\bf u}_{1}={\bf e}%
_{\alpha }\mbox{ such that }\alpha =\arg \max_{j}|b_{1j}|\ =\arg \max_{j}||%
{\bf A}_{\ast j}||\mbox{ }.
\]

c) The taxicab decomposition is based on $||{\bf A}||_{\infty \rightarrow
1}=||{\bf A}^{\prime }||_{\infty \rightarrow 1};$ its transition formulas are%
\[
{\bf Au}_{1}={\bf a}_{1}\mbox{ \ and \ }{\bf v}_{1}=sgn({\bf a}_{1})
\]%
and%
\[
{\bf A}^{\prime }{\bf v}_{1}{\bf =b}_{1}\mbox{ \ and \ }{\bf u}_{1}=sgn({\bf %
b}_{1}).
\]%
It is the most robust among all the transposition invariant induced norms
considered in this paper, and is used extensively by Choulakian and
coworkers in developing taxicab correspondence analysis: Choulakian (2004,
2006, 2008a, 2008b, 2013, 2014), Choulakian et al. (2006, 2013a, 2013b,
2014). We also note that the taxicab decomposition of a covariance matrix is
equivalent to the centroid decomposition of the centred dataset. The taxicab
norm was first considered by Grothendieck, see the interesting story of the
Grothhendieck theorem and its many versions by Pisier (2012). Here, we cite
this remarkable result

{\bf Grothendieck Inequality: }Let ${\bf A=(}a_{ij})$ be a real matrix of
size $m\times n;$ then for $i=1,..,m$ and $j=1,...,n$%
\begin{eqnarray}
||{\bf A}||_{\infty \rightarrow 1}
&=&\max_{s_{i},t_{j}}\sum_{i=1}^{m}\sum_{j=1}^{n}a_{ij}s_{i}t_{j}\mbox{ \
subject to }(s_{i},t_{j})\in \{-1,1\}^{2}  \nonumber \\
&=&\max_{s_{i},t_{j}}\sum_{i=1}^{m}\sum_{j=1}^{n}a_{ij}<s_{i},t_{j}>\mbox{%
subject to }(s_{i},t_{j})\in S_{2}^{1}\times S_{2}^{1}  \nonumber \\
&\leq &K_{d}\max_{{\bf s}_{i},{\bf t}_{j}}\sum_{i=1}^{m}\sum_{j=1}^{n}a_{ij}<%
{\bf s}_{i},{\bf t}_{j}>\mbox{subject to }({\bf s}_{i},{\bf t}_{j})\in
S_{2}^{d}\times S_{2}^{d},
\end{eqnarray}%
where $K_{d}$ is a universal constant that depends on $d$ for $d=2,3,...,$
but does not depend on $m$ and $n$. By defining $K_{1}=1,$ we see that $%
K_{d} $ $\geq K_{d-1}$ for $d=2,3,....$ The open problem is that there
exists a universal constant $K_{G}$ such that
\[
K_{G}=\inf_{d}K_{d}\mbox{ such that the inequality in (12) is true.}
\]%
It is conjectured that
\[
1.67695\leq K_{G}\leq \frac{\pi }{2\log (1+\sqrt{2})}=1.78221.
\]

An elementary proof of the inequality is given by Blei (1987) or Jameson
(1987). A randomization algorithm to compute $||{\bf A}||_{\infty
\rightarrow 1}$ via the Grothendieck inequality is studied by Alon and Naor
(2006), and it is used by McCoy and Tropp (2011) to compute (11). Rohn
(2000) shows that the computation of $||{\bf A}||_{\infty \rightarrow 1}$ is
NP-hard.

d) The singular value decomposition, SVD, is the standard decomposition, the
most used and studied; it is based on $||{\bf A}||_{2\rightarrow 2}=||{\bf A}%
^{\prime }||_{2\rightarrow 2}$, see \ Horn and Johnson (1990) and Golub and
Van Loan (1996). Its transition formulas are%
\[
{\bf Au}_{1}={\bf a}_{1}\mbox{ \ and \ }{\bf v}_{1}={\bf a}_{1}/\sqrt{{\bf a}%
_{1}^{\prime }{\bf a}_{1}}
\]%
and%
\[
{\bf A}^{\prime }{\bf v}_{1}{\bf =b}_{1}\mbox{ \ and \ }{\bf u}_{1}={\bf b}%
_{1}/\sqrt{{\bf b}_{1}^{\prime }{\bf b}_{1}}.
\]%
\bigskip

{\bf Example 2: }Let us compute a few decompositions to the following matrix%
\begin{equation}
\textbf{X}=\left(
  \begin{array}{cc}
    1 & -2 \\
    -2 & 4 \\
    0 & 2 \\
  \end{array}
\right)
\end{equation}

a) Taxicab decomposition: $||{\bf X}||_{\infty \rightarrow 1}$ is attained
at one of the axes: ${\bf u}^{\prime }=(1\ \ \ 1)\ \ $or \ $(1\ -1)$. For $%
{\bf u}^{\prime }=(1\ \ \ 1)$, $({\bf Xu})^{\prime }=(-1\ \ 2\ \ 2),\ $and
$||{\bf Xu}||_{1}=5.$ For ${\bf u}^{\prime
}=(1\ \ \ -1)$, $({\bf Xu})^{\prime }=(3\ \ -6\ \ -2),\ $and $||{\bf Xu}||_{1}=11.\ $So, ${\bf u}_{1}^{\prime }=(1\
\ \ -1),$\ ${\bf a}_{1}^{\prime }=(3\ \ -6\ \ -2),$ ${\bf v}_{1}^{\prime
}=sgn({\bf a}_{1}^{\prime })=(1\ \ -1\ \ -1),$ ${\bf b}_{1}^{\prime }=({\bf X%
}^{\prime }{\bf v}_{1})^{\prime }=(3\ \ -8),$ $\lambda _{1}=11=||{\bf a}%
_{1}||_{1}={\bf a}_{1}^{\prime }{\bf v}_{1}=||{\bf b}_{1}||_{1}={\bf b}%
_{1}^{\prime }{\bf u}_{1}.$ Note that ${\bf u}_{1}=sgn({\bf b}_{1}).$ Now
the residual matrix, ${\bf X}^{(1)}{\bf =X-a}_{1}{\bf b}_{1}^{\prime
}/\lambda _{1},$ is%
\[
{\bf X}^{(1)}=
\left(
  \begin{array}{cc}
    2 & 2 \\
    -4 & -4 \\
    6 & 6 \\
  \end{array}
\right)/11,
\]%
which is of rank 1. Repeating the above calculations on ${\bf X}^{(1)}$, we
find ${\bf u}_{2}^{\prime }=(1\ \ \ 1),$\ ${\bf a}_{2}^{\prime }=(1\ \ -2\ \
3)\ 4/11,$ ${\bf v}_{2}^{\prime }=sgn({\bf a}_{2}^{\prime })=(1\ \ -1\ \ 1),$
${\bf b}_{2}^{\prime }=({\bf X}^{\prime }{\bf v}_{2})^{\prime }=(1\ \ 1)\
12/11,$ $\lambda _{2}=24/11=||{\bf a}_{2}||_{1}={\bf a}_{2}^{\prime }{\bf v}%
_{2}=||{\bf b}_{2}||_{1}={\bf b}_{2}^{\prime }{\bf u}_{2}.$ Note that ${\bf u%
}_{2}=sgn({\bf b}_{2}).$ Now the residual matrix, ${\bf X}^{(2)}{\bf =X}%
^{(1)}{\bf -a}_{2}{\bf b}_{2}^{\prime }/\lambda _{2}={\bf 0.}$ So we have
the following decomposition%
\begin{equation}
{\bf X\ }{\bf =\ }(3\ \ -6\ \ -2)^{\prime }(3\ \ -8)/11+(1\ \ -2\ \
3)^{\prime }(1\ \ 1)2/11.
\end{equation}

b) Centroid decomposition: $||{\bf X}||_{\infty \rightarrow 2}$ is attained
at one of the axes: ${\bf u}^{\prime }=(1\ \ \ 1)\ \ $or \ $(1\ -1)$. For $%
{\bf u}^{\prime }=(1\ \ \ 1)$, $({\bf Xu})^{\prime }=(-1\ \ 2\ \ 2),\ $and
$||{\bf Xu}||_{2}=\sqrt{9}.$ For ${\bf u}%
^{\prime }=(1\ \ \ -1)$, $({\bf Xu})^{\prime }=(3\ \ -6\ \ -2),\ $and
$||{\bf Xu}||_{2}=7.\ $So, ${\bf u}%
_{1}^{\prime }=(1\ \ \ -1),$\ ${\bf a}_{1}^{\prime }=(3\ \ -6\ \ -2),$ ${\bf %
v}_{1}^{\prime }={\bf a}_{1}^{\prime }/\lambda _{1}=(3\ \ -6\ \ -2)/7,$ $%
{\bf b}_{1}^{\prime }=({\bf X}^{\prime }{\bf v}_{1})^{\prime }=(15\ \
-34)/7, $ $\lambda _{1}=7=||{\bf a}_{1}||_{2}={\bf a}_{1}^{\prime }{\bf v}%
_{1}=||{\bf b}_{1}||_{1}={\bf b}_{1}^{\prime }{\bf u}_{1}.$ Note that ${\bf u%
}_{1}=sgn({\bf b}_{1}).$ Now the residual matrix, ${\bf X}^{(1)}{\bf =X-a}%
_{1}{\bf b}_{1}^{\prime }/\lambda _{1},$ is%
\[
{\bf X}^{(1)}=\left(
                \begin{array}{cc}
                  4 & 4 \\
                  -8 & -8 \\
                  30 & 30 \\
                \end{array}
              \right)/49,
\]%
which is of rank 1. Repeating the above calculations on ${\bf X}^{(1)}$, we
find ${\bf u}_{2}^{\prime }=(1\ \ \ 1),$\ ${\bf a}_{2}^{\prime }=(2\ \ -4\ \
15)\ 4/49,$ ${\bf v}_{2}^{\prime }={\bf a}_{2}^{\prime }/||{\bf a}%
_{2}||_{2}=(2\ \ -4\ \ 15)\ /(7\sqrt{5}),$ ${\bf b}_{2}^{\prime }=({\bf X}%
^{\prime }{\bf v}_{2})^{\prime }=(1\ \ 1)\ 2\sqrt{5}/7,$ $\lambda _{2}=4%
\sqrt{5}/7=||{\bf a}_{2}||_{2}={\bf a}_{2}^{\prime }{\bf v}_{2}=||{\bf b}%
_{2}||_{1}={\bf b}_{2}^{\prime }{\bf u}_{2}.$ Note that ${\bf u}_{2}=sgn(%
{\bf b}_{2}).$ Now the residual matrix, ${\bf X}^{(2)}{\bf =X}^{(1)}{\bf -a}%
_{2}{\bf b}_{2}^{\prime }/\lambda _{2}={\bf 0.}$ So we have the following
decomposition%
\begin{equation}
{\bf X=\ }(3\ \ -6\ \ -2)^{\prime }(15\ \ -34)/49+(2\ \ -4\ \ 15)^{\prime
}(1\ \ 1)2/49.
\end{equation}

c) Extreme decomposition: $||{\bf X}||_{1\rightarrow \infty }$ is attained
on one of the canonical basis vectors: ${\bf e}_{1}^{\prime }=(1\ \ \ 0)\ \ $%
or \ ${\bf e}_{2}^{\prime }=(0\ \ \ 1)$. For ${\bf u=e}_{1}$, $({\bf Xu}%
)^{\prime }=(1\ \ -2\ \ 0),\ $and ${||\bf %
Xu}||_{\infty }=2.$ For ${\bf u=e}_{2}$, $({\bf Xu})^{\prime }=(-2\ \ 4\ \
2),\ $and $||{\bf Xu}||_{\infty }=4.\ $So,
${\bf u}_{1}={\bf e}_{2},$\ ${\bf a}_{1}={\bf X}_{\ast 2}$ the second column
of ${\bf X},$ ${\bf v}_{1}^{\prime }=(0\ \ 1\ \ 0),$ ${\bf b}_{1}^{\prime }=(%
{\bf X}^{\prime }{\bf v}_{1})^{\prime }=(-2\ \ 4),$ $\lambda _{1}=4=||{\bf a}%
_{1}||_{\infty }={\bf a}_{1}^{\prime }{\bf v}_{1}=||{\bf b}_{1}||_{\infty }=%
{\bf b}_{1}^{\prime }{\bf u}_{1}.$ Note that ${\bf u}_{1}={\bf e}_{2}$ $%
sgn(b_{12}).$ Now the residual matrix, ${\bf X}^{(1)}{\bf =X-a}_{1}{\bf b}%
_{1}^{\prime }/\lambda _{1},$ is%
\[
{\bf X}^{(1)}=\left(
                \begin{array}{cc}
                  0 & 0 \\
                  0 & 0 \\
                  1 & 0 \\
                \end{array}
              \right)/11,
\]%
which is of rank 1. Repeating the above calculations on ${\bf X}^{(1)}$, we
find ${\bf u}_{2}={\bf e}_{1},$\ ${\bf a}_{2}^{\prime }=(0\ \ 0\ \ 1),$ $%
{\bf v}_{2}={\bf a}_{2},$ ${\bf b}_{2}^{\prime }=({\bf X}^{\prime }{\bf v}%
_{2})^{\prime }=(1\ \ 0),$ $\lambda _{2}=1=||{\bf a}_{2}||_{\infty }={\bf a}%
_{2}^{\prime }{\bf v}_{2}=||{\bf b}_{2}||_{\infty }={\bf b}_{2}^{\prime }%
{\bf u}_{2}.$ Note that ${\bf u}_{2}={\bf e}_{1}sgn(b_{21}).$ Now the
residual matrix, ${\bf X}^{(2)}{\bf =X}^{(1)}{\bf -a}_{2}{\bf b}_{2}^{\prime
}/\lambda _{2}={\bf 0.}$ So we have the following decomposition%
\begin{equation}
{\bf X=\ (}-2\ \ 4\ \ 2{\bf )}^{\prime }(-2\ \ 4)/4+(0\ \ 0\ \ 1)^{\prime
}(1\ \ 0).
\end{equation}

d) Singular value decomposition: It is based on the eigenvalues and
eigenvectors of the symmetric matrix ${\bf X}^{\prime }{\bf X}$: $\lambda
_{1}=5.3191,$ ${\bf v}_{1}^{\prime }=(-0.4197\ \ 0.8393\ \ 0.3455),\ {\bf a}%
_{1}=\lambda _{1}{\bf v}_{1},$\ ${\bf u}_{1}=(-0.3945\ \ 0.9189)$ and ${\bf b%
}_{1}=\lambda _{1}{\bf u}_{1};\ \lambda _{2}=0.8408,\ {\bf v}_{2}^{\prime
}=(-0.1545\ \ 0.3090\ \ -0.9384),$ ${\bf a}_{1}=\lambda _{1}{\bf v}_{1},$\ $%
{\bf u}_{2}=(-0.9189\ \ -0.3945)$ and ${\bf b}_{2}=\lambda _{2}{\bf u}_{2}.\
$So we have the following decomposition
\begin{eqnarray}
{\bf X} &{\bf =}&\ (-0.4197\ \ 0.8393\ \ 0.3455)^{\prime }(-0.3945\ \
0.9189)5.3191+  \nonumber \\
&&(-0.1545\ \ 0.3090\ \ -0.9384)^{\prime }(-0.9189\ \ -0.3945)0.8408.
\end{eqnarray}

{\bf Remark 2:}

a) We note that the factors $({\bf a}_{\alpha },{\bf b}_{\alpha })$ are
determined up to proportionality, and the four decompositions in equations
(14) through (17) of the data set X given in (13) are essentially different.
This is a much discussed and important topic, named factor indeterminacy
problem; see for instance Mulaik (1987). We can recast or reformulate the
factor indeterminacy problem within a geometric setting: If $Rank({\bf X}%
)\geq 2,$ then there are infinite number of different factorizations
depending on the values of $r\geq 1$ and $p\geq 1$ for ${\bf X}\in
B(l_{r}^{n},l_{p}^{m}).$

b) The decomposition of ${\bf X}$ is essentially unique (up to
proportionality) if and only if $rank({\bf X})=1$.

c) Conditions for essential uniqueness of decompositions for three-way
arrays or tensors is an active area of research; and Kruskal's sufficiency
theorem is the most famous general result, see Rhodes (2010). For an
overview of the literature on tensor decomposition, see the interesting
review by Ten Berge (2011).

\subsection{A\ family\ of\ Euclidean\ multidimensional\ scaling\ models}

Let ${\bf \Delta }=(\delta _{ij})$ be\ a symmetric $n\times n$ matrix with
nonnegative elements and zeros on the diagonal, representing the
dissimilarities of $n$ objects. The aim of multidimensional scaling (MDS)
techniques is to find a configuration of the $n$ points, which best match
the original dissimilarities as much as possible. We shall consider the
framework of the classical MDS, named also principle coordinate analysis,
where we suppose that the dissimilarities represent Euclidean distances,
which implies that there exists a set of n centered points in a Euclidean
space, denoted by $\{{\bf f}_{i}:i=1,...,n\},$ such that
\[
\delta _{ij}^{2}=\left\vert \left\vert {\bf f}_{i}-{\bf f}_{j}\right\vert
\right\vert _{2}^{2}.
\]%
We thus have the following well known relationship%
\begin{eqnarray*}
{\bf Q} &=&-\frac{1}{2}({\bf H\Delta }^{2}{\bf H}) \\
&=&{\bf F}^{\prime }{\bf F},
\end{eqnarray*}%
where ${\bf F=[f}_{1},...,{\bf f}_{n}]$ and ${\bf H}={\bf I}_{n}-{\bf 1}_{n}%
{\bf 1}_{n}^{\prime }/n$ is the centering matrix, ${\bf I}_{n}$ the identity
matrix and ${\bf 1}_{n}$ the vector of ones. So the matrix ${\bf Q}$ is
positive semi-definite, and it is equal to ${\bf F}^{\prime }{\bf F}$, where
${\bf F}$ is unknown. Now suppose that ${\bf F\in \ }B(l_{r}^{n},l_{p}^{m}).$
Then
\[
||{\bf F}||_{r\rightarrow p}=\max \{||{\bf Fu}||_{p}:{\bf u}\in S_{r}^{n}\}
\]%
can be expressed as a linear finction of ${\bf Q}$ if $p=2$ and $r\geq 1;$
that is
\begin{eqnarray}
||{\bf F}||_{r\rightarrow 2}^{2} &=&{\bf u}^{\prime }{\bf F}^{\prime }{\bf Fu%
}\mbox{ \ \ subject to }{\bf u}\in S_{r}^{n}  \nonumber \\
&=&{\bf u}^{\prime }{\bf Qu}\mbox{ \ \ subject to }{\bf u}\in S_{r}^{n}.
\end{eqnarray}%
Factorizing ${\bf Q}$ into ${\bf F}^{\prime }{\bf F}$ by (18), we obtain a
family of Euclidean multidimensional scaling (MDS) models as a function of $%
r\geq 1$. Three particular cases are worthy of mention:

a) For $r=2$, we obtain the classical MDS, see for instance Torgerson (1952)
and Gower (1966). $\ $Each ${\bf f}_{\alpha }$ is an eigenvector of ${\bf Q}$%
; see also subsection 2.1 part c).

b) For $r=\infty ,$ we get the centroid MDS, where we maximize the
Burt-Thurstone criterion (11), see also subsection 2.1 part a).

c) For $r=1,$ we get the dominant MDS; its computation is extremely simple
and fast, see subsection 2.1 part b).

\bigskip

{\bf Example 3: }We consider the Facial Expressions data found in Borg and
Groenen (2005, p. 76) of dimension 13x13, where $n=13$ is the number of
person's facial expressions. The aim of the study is the correct
identification of intended emotional message from a person's facial
expression. Furthermore, Table 4.3, p. 75 in Borg and Groenen, provide
Schlosberg empirical scale values that classify the facial expressions into
three classes: pleasant-unpleasant (PU), attention-rejection (AR) and
tension-sleep (TS). Borg and Groenen (2005, subsection 4.3) found that the
first two dimensions of ordinal MDS reproduced quite accurately the three
classes: the first dimension representing PU and the second dimension
representing AR and TS, because the correlations between the first two
calculated dimensions and the Schlosberg empirical scale values are quite
high for ordinal MDS. Table 1 compares the correlation values obtained by
four MDS approaches; the ordinal MDS correlation values are reproduced from
Borg and Groenen (2005, p.77, Table 4.6): The centroid MDS produced results
as good as the ordinal MDS. \bigskip

\begin{tabular}{|l||l|l|l|}
\multicolumn{4}{l||}{\bf Table 1: Facial Expressions Data: Correlation
values.} \\ \hline
\multicolumn{1}{|l|}{MDS} & \multicolumn{1}{||l|}{corr(DIM1,PU)} &
corr(DIM2,AR) & corr(DIM2,TS) \\ \hline
\multicolumn{1}{|l|}{ordinal} & \multicolumn{1}{||l|}{0.94} & 0.86 & 0.87 \\
\hline
\multicolumn{1}{|l|}{classical} & \multicolumn{1}{||l|}{0.91} & 0.80 & 0.83
\\ \hline
\multicolumn{1}{|l|}{centroid} & \multicolumn{1}{||l|}{0.93} & 0.86 & 0.89
\\ \hline
\multicolumn{1}{|l|}{dominant} & \multicolumn{1}{||l|}{0.91} & 0.78 & 0.70
\\ \hline
\end{tabular}

\section{The French school of data analysis}

Benz\'{e}cri (1973), who was a pure mathematician in geometry in the 1950s,
is considered the father of the french school of data analysis; he developed
a geometric generalized Euclidean framework for multidimensional data
analysis by introducing two metric matrices (square and positive definite) $%
{\bf M}$ and ${\bf N.}$ In this setting, the duality diagram of Theorem 1
becomes\bigskip

\ \ \ \ \ \ \ \ \ \ \ \ \ \ \ \ \ \ \ \ \ ${\bf A}$

\ \ \ \ \ \ \ \ $S_{2}^{n}({\bf N})\longrightarrow \ \ l_{2}^{m}$

$\ {\bf N}=\varphi \uparrow \ \ \ \ \ \ \ \ \ \ \ \downarrow \varphi ={\bf M}
$

\ \ \ \ \ \ \ \ \ \ \ \ $l_{2}^{n}\ \ \longleftarrow \ \ S_{2}^{m}({\bf M})\
$

\ \ \ \ \ \ \ \ \ \ \ \ \ \ \ \ \ \ \ $\ {\bf A}^{\prime }\bigskip $

where
\[
S_{2}^{n}({\bf N})=\{{\bf x}\in
\mathbf{R}
\ ^{\mbox{n}}:{\bf x}^{\prime }{\bf Nx}=\mbox{1}\},
\]%
and
\[
S_{2}^{m}({\bf M})=\{{\bf x}\in
\textbf{R}
\ ^{\mbox{m}}:{\bf x}^{\prime }{\bf Mx}=\mbox{1}\}.
\]%
Note that, for ${\bf N=I}_{n}$, then $S_{2}^{n}({\bf I}_{n})=S_{2}^{n}.$ In
this particular Euclidean setting, the duality diagram represents the
following transition fomulas%
\[
{\bf Au}_{1}={\bf a}_{1}\mbox{ \ and \ }{\bf v}_{1}={\bf Ma}_{1}/\sqrt{{\bf a%
}_{1}^{\prime }{\bf Ma}_{1}})
\]%
and%
\[
{\bf A}^{\prime }{\bf v}_{1}{\bf =b}_{1}\mbox{ \ and \ }{\bf u}_{1}={\bf Nb}%
_{1}/\sqrt{{\bf b}_{1}^{\prime }{\bf Nb}_{1}}).
\]

The solution of the last two equations can be reexpressed as a generalized
eigenvalue-eigenvector problem in the following way: From the last two
equations we get
\begin{eqnarray*}
{\bf a}_{1} &=&{\bf Au}_{1} \\
&=&{\bf ANb}_{1}/\sqrt{{\bf b}_{1}^{\prime }{\bf Nb}_{1}}) \\
{\bf M}^{-1}{\bf v}_{1}\sqrt{{\bf a}_{1}^{\prime }{\bf Ma}_{1}}) &=&{\bf ANA}%
^{\prime }{\bf v}_{1}/\sqrt{{\bf b}_{1}^{\prime }{\bf Nb}_{1}}),
\end{eqnarray*}%
from which one gets%
\[
{\bf MANA}^{\prime }{\bf v}_{1}=\lambda _{1}{\bf v}_{1};
\]%
and similarly%
\[
{\bf NA}^{\prime }{\bf MAu}_{1}=\lambda _{1}{\bf u}_{1}.
\]%
In the last two equations the eigenequations are functions of principal axes
${\bf u}_{1}$ and ${\bf v}_{1}.$ However one can reexpress them as functions
of projected factor scores%
\[
{\bf ANA}^{\prime }{\bf Ma}_{1}=\lambda _{1}{\bf a}_{1};
\]%
and similarly%
\[
{\bf NA}^{\prime }{\bf MANb}_{1}=\lambda _{1}{\bf b}_{1}.
\]

\section{\qquad Conclusion}

We embedded the ordinary SVD into a larger family based on induced matrix
norms, and provided the transition formulas and a simple criss-cross
iterative procedure to compute the principal axes and principal factor
scores. Given that there are infinite number of SVD like decompositions,
depending on the underlying induced norms, one is tempted to ask which is
the best?

It is quite ironic that the centroid decomposition, the oldest method, was
recently rediscovered and restudied as a robust method, see Choulakian
(2005), Kwak (2008), McCoy and Tropp (2011), after being dumped almost sixty
years ago for the following reason given in Hubert et al. (2000, p.76)
"Comments in Guttman (1944) and elsewhere (e.g., Horst (1965) and Harman
(1967)) with regard to this centroid strategy generally considered it a poor
approximation to what could be generated from Hotelling's method that would
choose successive unit length vectors to produce a rank reduction by
identifying (through an iterative strategy) the eigenvector associated with
the largest eigenvalue for each of the residual matrices successively
obtained. At the time, however, the centroid method was computationally much
less demanding than Hotelling's iterative (or power) method for obtaining
each of the principal components (again, one-at-a-time and reducing rank at
each iteration); for this reason alone, the centroid method was a very
common factorization strategy until electronic computing capabilities became
more widely available". This comment shows that the centroid method was a
victim of the habit of using mathematical methods in statistics based on
optimal criteria, as if optimality is a guarantee of efficiency.

The arguments advanced by Benz\'{e}cri on the advantages of the Euclidean
geometry over the taxicab geometry for multidimensional data analysis are
both computational and metaphysical. On the use of L1 distance in data
analysis, Benz\'{e}cri (1977, page 13) commented in the following way " elle
(L1) ne permet pas d'utiliser la g\'{e}om\'{e}trie euclidienne
multidimensionnelle; elle donnera des r\'{e}sultats qui qualitativement
ressembleront \`{a} ceux obtenus par la distance...quadratique; mais au prix
de calculs plus compliqu\'{e}s et sous une forme moins commode. Sans
permettre \`{a} l'outil math\'{e}matique de d\'{e}figurer le r\'{e}el, on
doit lui conc\'{e}der que la transmission \`{a} l'esprit humain d'un vaste
ensemble de donn\'{e}es synth\'{e}tis\'{e} (r\'{e}sum\'{e}; rendu
perceptible par le calcul) ait ses lois propres. (On se souvient que le
primat de la g\'{e}om\'{e}trie euclidienne est admis par Torgerson)". The
ease of computation argument is very similar to Gauss's argument in the
adoption of least squares criterion in the linear regression model. While
the metaphysical argument, if my understanding is correct, is that: the
transmission to the human spirit of a synthesis of a collection of data has
its proper laws, which are based on the Euclidean geometry.

Ten Berge (2005, personal communication) thought that the centroid method
produced good results, but it was not mathematically well understood during
the last century. Benz\'{e}cri (1973b, page 1 in Avant-propos) considered
data analysis an experimental science, and that he has a predeliction for
the case studies in his books. A similar thought is also found in Tukey:
"The test of a good procedure is how well it works, not how well it is
understood".

As to the question asked which decomposition is the best? Mathematically,
the SVD is the best and the reference, but quite sensitive to outlying
observations: So, we suggest the joint use of SVD and the taxicab
decomposition, or, the SVD and the centroid decomposition. \bigskip

{\em Acknowledgement: }This research was financed by the Natural Sciences
and Engineering Research Council of Canada. The author thanks Bernard
Fichet, Fran\c{c}ois Drucker and Pascal Pr\'{e}a of the \'{E}cole Centrale
de Marseille for the invitation, for which this paper has been specially
prepared.\bigskip\

{\bf References}\bigskip

Alon, N. and Naor, A. (2006). Approximating the cut-norm via Grothendieck's
inequality.\ {\it SIAM Journal on Computing}, 35, 787-803.

Benz\'{e}cri, J.P. (1973a).\ {\it L'Analyse des Donn\'{e}es: Vol. 2:
L'Analyse des Correspondances}. Paris: Dunod.

Benz\'{e}cri, J.P. (1973b).\ {\it L'Analyse des Donn\'{e}es: Vol. 1: La
Taxinomie}. Paris: Dunod.

Benz\'{e}cri, J.P. (1973). Histoire et Pr\'{e}histoire de l'Analyse des Donn%
\'{e}es: V: L'Analyse des correspondances. Les Cahiers de l'Analyse des Donn%
\'{e}es, II(1), 9-53.

Blei, R.C. (1987). An elementary proof of the Grothendieck inequality. {\it %
Proceedings of the American Mathematical Society}, 100(1), 58-60.

Borg, I. and Groenen, P.G. (2005). {\it Modern Multidimensional Scaling}.
2nd edition, NY: Springer Verlag.

Boyd, D.W. (1974). The power method for $l_{p}$ norms. {\it Linear Algebra
and its Applications}, 9, 95-101.

Burt, C. (1917). {\it The Distribution and Relations of Educational Abilities%
}. London, U.K: P.S. King \& son.

Choulakian, V. (2003). The optimality of the centroid method. {\it %
Psychometrika}, 68, 473-475.

Choulakian, V. (2004). A comparison of two methods of principal component
analysis. In {\it COMPSTAT'2004} edited by J. Antoch,
Physica-Verlag/Springer, 793-798.

Choulakian, V. (2005a). Transposition invariant principal component analysis
in L$_{1}$ for long tailed data. {\it Statistics and Probability Letters},
71, 23-31.

Choulakian, V. (2005b). L$_{1}$-norm projection pursuit based principal
component analysis. {\it Computational Statistics and Data Analysis}, 50,
1441-1451.

Choulakian, V. (2006a). Taxicab correspondence analysis. {\it Psychometrika,}
71, 333-345.

Choulakian, V. (2006b). L$_{1}$ norm projection pursuit principal component
analysis. {\it Computational Statistics and Data Analysis}, 50, 1441-1451.

Choulakian, V. (2008a). Taxicab correspondence analysis of contingency
tables with one heavyweight column. {\it Psychometrika}, 73, 309-319.

Choulakian, V. (2008b). Multiple taxicab correspondence analysis. {\it %
Advances in data Analysis and Classification}, 2, 177-206.

Choulakian, V. (2012). Picture of all solutions of successive 2-block MAXBET
problems. {\it Psychometrika}, 76(4), 550-563.

Choulakian V. (2013). The simple sum score statistic in taxicab
correspondence analysis. In {\it Advances in Latent Variables }(ebook), eds.
Brentari E.and Carpita M., Vita e Pensiero, Milan, Italy, ISBN 978 88 343
2556 8, 6 pages.

Choulakian, V. (2014). Taxicab correspondence analysis of ratings and
rankings. Journal de la Soci\'{e}t\'{e} Fran\c{c}aise de Statistique.
155(4), 1-23.

Choulakian, V., Allard, J. and Simonetti, B. (2013). Multiple taxicab
correspondence analysis of a survey related to health services. {\it Journal
of Data Science}, 11(2), 205-229.

Choulakian, V. and de Tibeiro, J. (2013). Graph partitioning by
correspondence analysis and taxicab correspondence analysis. {\it Journal of
Classification}, 30(3), 397-427.

Choulakian, V., Simonetti, B. and Gia, T.P. (2014). Some new aspects of
taxicab correspondence analysis. {\it Statistical Methods and Applications},
23(3), 401-416.

Choulakian, V., Kasparian, S., Miyake, M., Akama, H., Makoshi, N., Nakagawa,
M. (2006). A statistical analysis of synoptic gospels. {\it JADT'2006}, pp.
281-288.

Chu, M.T. and Funderlic, R.E. (2002). The centroid decomposition:
relationships between discrete variational decompositions and SVDs. {\it %
Siam J. Matrix Analysis and Applications}, 23, 1025-1044.

Chu, M.T., Funderlic, R.E. and Golub, G.H. (1995). A rank-one reduction
formula and its applications to matrix factorizations. {\it SIAM Review},
37(4), 512-530.

De La Cruz, O. and Holmes, S. (2011). The duality diagram in data analysis:
Examples of modern applications. {\it Annals of Applied Statistics}, 5(4),
2266-2277.

Drakakis. K. and Pearlmutter, B.A. (2009). On the calculation of the $%
l_{2}\rightarrow l_{1}$ niduced matrix norm. {\it International Journal of
Algebra}, 3(5), 231-240.

Gabriel, K.R. and Zamir, S. (1979). Lower rank approximation of matrices by
least squares with any choice of weights. {\it Technometrics,} 21, 489-498.

Galpin, J.S. and Hawkins, D.M. (1987). L$_{1}$ estimation of a covariance
matrix. {\it Computational Statistics and Data Analysis}, 5, 305-319.

Golub, G.H. and Van Loan, C.F. (1996). {\it Matrix Computations}. 3rd
edition, John Hopkins Studies in the Mathematical Sciences, The Johns
Hopkins University Press, Baltimore, MD, ISBN 978-0-8018-5414-9.

Gower, J.C. (1966). Some distance properties of latent root and vector
methods in multivariate analysis. {\it Biometrika}, 53, 325-338.

Guttman, L. (1944). General theory and methods for matric factoring.
Psychometrika, 9, 1-16.

Harman, H.H. (1967). {\it Modern Factor Analysis}. Chicago, IL: The
University of Chicago Press.

Holmes, S. (2008). Mutivariate analysis: The French way. In {\it Probability
and Statistics: Essays in honor of David Freedman}, eds. D. Nolan and T.
Speed, pp. 219-233. Ohio, USA: Institute of Mathematical Statistics.

Horn, R.A. and Johnson, C.R. (1990). {\it Matrix Analysis}. Cambridge
University Press, NY.

Horst, P. (1965). {\it Factor Analysis of Data Matrices}. NY: Holt, Rinehart
and Winston.

Hubert, L., Meulman, J. and Heiser, W. (2000). Two purposes for matrix
factorization: A historical appraisal. {\it SIAM Review}, 42, 68-82.

Jameson, G.J.O. (1987). {\it Summing and Nuclear Norms in Banach Space Theory%
}. London Mathematical Society Student Texts 8, Cambridge University Press,
Cambridge.

Kreyszig, E. (1978). {\it Introductory Functional Analysis with Applications}%
. J. Wiley and Sons, N.Y.

Kwak, N. (2008). Principal component analysis based on L1-norm maximization.
{\it IEEE Transactions on Pattern Analysis and Machine Intelligence}, 30(9),
1672-1680.

Lewis, A.D. (2010). A top nine list: Most popular induced matrix norms.
Downloaded from www.mast.queensu.ca/\symbol{126}andrew/notes/pdf/2010b.pdf

McCoy, M. and Tropp, J. (2011). Two proposals for robust PCA using
semidefinite programming. {\it Electronic Journal of Statistics}, 5,
1123-1160.

Mulaik, S. (1987). A brief history of the phlosophical foundations of
exploratory factor analysis. {\it Multivariate Behavioral Research}, 22,
267-305.

Rhodes, J.A. (2010). A concise proof of Kruskal's theorem on tensor
decomposition. {\it Linear Algebra and its Applications}, 432 (7), 1818-1824.

Ten Berge, J.M.F. (2011). Simplicity and typical rank results for three-way
arrays. {\it Psychometrika}, 76(1), 3-12.

Thurstone, L.L. (1931). Multiple factor analysis. {\it Psychological Review}%
, 38, 406-427.

Thurstone, L.L. (1947). {\it Multiple-Factor Analysis}. The University of
Chicago Press, Chicago.

Torgerson, W.S. (1952). Multidimensional scaling: 1. Theory and method. {\it %
Psychometrika}, 17, 401-419.

Wold, H. (1966). Estimation of principal components and related models by
iterative least squares. In {\it Multivariate Analysis, }ed. Krishnaiah,
P.R., N.Y: Academic Press, 391-420.

\end{document}